\documentclass[12pt]{iopart}
\usepackage{graphicx}
\usepackage{bm}
\usepackage{subfigure}
\usepackage{cite}

\begin{document}
\title{Entanglement, Subsystem Particle Numbers and Topology in Free Fermion Systems}
\author{Y. F. Zhang$^1$, L. Sheng$^1$, R. Shen$^1$,Rui Wang $^1$ $^2$, D. Y. Xing$^1$}
\address{$^1$National Laboratory of Solid State Microstructures and
Department of Physics, Nanjing University, Nanjing 210093, China\\
$^2$Department of Physics, Zhejiang Ocean University, Zhoushan 316000, China}
\ead{shengli@nju.edu.cn}

\begin{abstract}
We study the relationship between bipartite entanglement,
subsystem particle number and topology in a half-filled free fermion system. It is proposed that
the spin-projected particle numbers can distinguish the quantum spin
Hall state from other states, and can be used to establish a new topological
index for the system. Furthermore, we apply the new topological invariant
to disordered system and show that a topological phase transition occurs when the
disorder strength is increased beyond a critical values. It is also shown that the subsystem
particle number fluctuation displays behavior very similar to the
entanglement entropy. It provides a lower-bound estimation for the
entanglement entropy, which can be utilized to obtain an
estimate of the entanglement entropy experimentally.
\end{abstract}
\maketitle

\section{Introduction}
Topological phases of matter are usually distinguished by using some
global topological properties, such as topological invariants and
topologically protected gapless edge modes, rather than certain
local order parameters. The integer quantum Hall effect~\cite{iqhe},
fractional quantum Hall effect~\cite{fqhe}, and band Chern
insulators~\cite{ci} can be characterized by Chern numbers or Berry
phases~\cite{tknn,tknn1}. The quantum spin Hall (QSH)
effect~\cite{qshe1,qshe2} and the three-dimensional topological
insulators~\cite{3d1,3d2} are characterized by the $Z_{2}$
invariant~\cite{z2} or spin Chern number~\cite{spinch1,spinch2}.
 In recent years, quantum entanglement~\cite{entangle1}, which reveals the
phase information of the quantum-mechanical ground-state
wavefunction, has been used as a tool to characterize the
topological phases. As shown by Levin and Wen~\cite{wen} and also by
Kitaev and Preskill~\cite{kitaev}, the existence of topological
entanglement entropy  in a fully gapped system, such as fractional
quantum Hall~\cite{teefqh} and the  gapped $Z_{2}$ spin liquid~\cite{wen,teeqsl}, indicates existence of long-range
quantum entanglement (topological order~\cite{toporder} in
equivalent parlance). Interestingly, a very recent  work~\cite{geoent} proved that topological order can also be read and assessed
by the geometric entanglement in multipartite entangled systems.
Another important progress is the
demonstration that the entanglement spectrum (ES)~\cite{es1} reveals
the gapless edge spectrum for fractional quantum Hall systems~\cite{es1,es2,es3,es4,es5}, Chern
insulators~\cite{esc1,esc2}, topological insulators~\cite{est1,est2,est3} and even to spin systems~\cite{ess1,ess2,ess3}.

Supposing $A$ and $B$ to be two blocks of a large system in a pure
quantum state, the reduced density matrix (RDM) $\rho_{A}$ can be
obtained by tracing over degrees of freedom of $B$. Then the Von
Neumann entanglement entropy (EE) can be computed
\begin{equation}
 S_{ent}=-\mbox{tr}(\rho_{A}\ln\rho_{A})=-\mbox{tr}(\rho_{B}\ln\rho_{B}) \ .
\end{equation}
It has been shown that for bipartite subsystems $A$ and $B$ with a
smooth boundary, $S_{ent}$ has  the form of $S_{ent}=\alpha
L-S_{top}$, where $L$ is the length of the boundary, $\alpha$ is a
non universal coefficient, and $-S_{top}$ is a universal constant
called the \emph{topological entanglement
entropy}~\cite{wen,kitaev}. Moreover, if we write the RDM in the
form  of $\rho_{A}=\exp(-H_{ent})/Z$, where $Z$ is a normalization
constant, and $H_{ent}$ is known as the \emph{entanglement
Hamiltonian}, the eigenvalue spectrum $\{\varepsilon_{i}\}$ of
$H_{ent}$ is called the ES, which stores more information about the
quantum entanglement than the EE~\cite{es1}.


\begin{figure}
  \centering
  \subfigure[\  cylinder geometry ]{
  \begin{minipage}[c]{0.4\linewidth}
    \centering
    \includegraphics[width=1.5in]{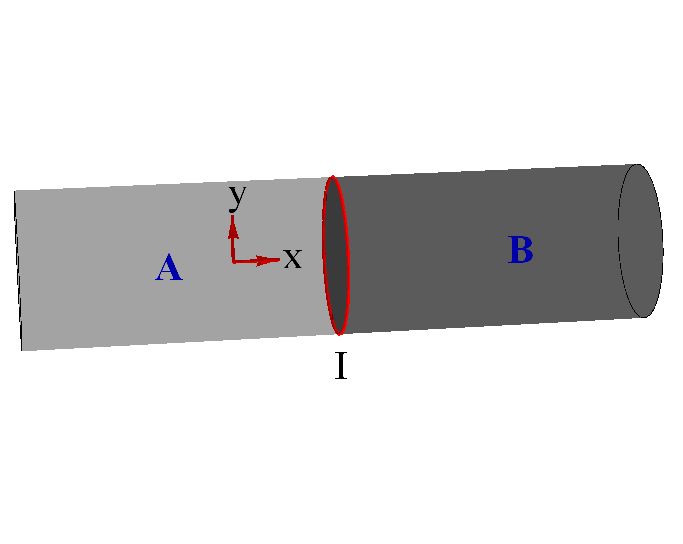}
  \end{minipage}%
  \label{fig:a}
  }%
  \subfigure[\  torus geometry]{
  \begin{minipage}[c]{0.4\linewidth}
    \centering
    \includegraphics[width=1.8in]{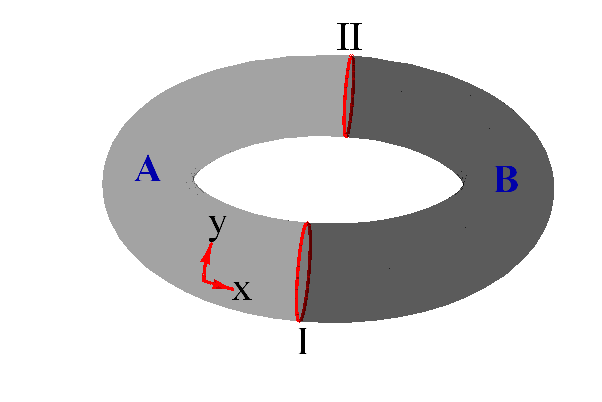}
  \end{minipage}
  \label{fig:b}
  }
\caption{(Color online) Schematic view of a cylinder and a torus. The entanglement
cuts divide the system into two equal parts $A$ and $B$. For the
cylinder geometry, the  entanglement cut leads to one interface (a);
and for the torus geometry, the cuts lead to two interfaces (b).}  
  \label{geo} 
\end{figure}

In this paper, we study the relationship between bipartite
entanglement and subsystem particle number in half-filled free
fermion systems.
It was proposed in Ref.~\cite{trace}, for systems with
translational invariance in one dimension, the discontinuity in the
subsystem particle number as a function of the conserved momentum
indicates whether or not the ES has a spectral flow, which is
determined by the topological invariant of the system~\cite{est3}. Nevertheless,
this approach has
an exceptional case for a half-filled QSH system
with two-dimensional inversion symmetry.
To overcome the inadequacy, we define spin-projected particle
numbers, based on which spin trace indices can be well
defined, for the QSH system with or without $s_{z}$ conservation.
Spin trace indices are univocally related to the topological invariant
of QSH system, i.e., the $Z_2$ index. Furthermore, we show that spin trace indices will still work well in
disordered systems and can demonstrate a topological phase transition. We further investigate the
relationship between the EE and subsystem particle number
fluctuation. The latter is also dominated by the boundary
excitations of the system, and satisfies a similar area law as the
EE.

In the next section, we introduce the model Hamiltonian,
and explain the procedure to calculate the ES and EE. In Sec.
III, numerical calculation of the ES is carried out, and
the connection between
the subsystem spin-projected particle numbers
and the topological invariants in different phases is established.
In Sec. IV, we apply
spin trace indices to disorder system and show it can demonstrate the
phase transition from the QSH  to the trivial
insulator.
 In Sec. V, the relationship between the EE and subsystem particle number
fluctuation is discussed. Our results are summarized and discussed in the final section.

\section{ MODEL HAMILTONIAN}
We begin with the tight-binding model Hamiltonian for the QSH system
introduced by Kane and Mele~\cite{qshe1,z2}, plus  an additional
exchange field~\cite{yang}
\begin{eqnarray}
 H&=-\sum_{\langle \bi{i},\bi{j}\rangle}c_{\bi{i}}^\dag
c_{\bi{j}}+ iv_{so}\sum\limits_{\langle\langle \bi{i},\bi{j}\rangle\rangle}c_{\bi{i}}^\dag
\sigma_{z}v_{ij}c_{\bi{j}} \nonumber\\
& +iv_{r}\sum\limits_{\langle \bi{i},\bi{j}\rangle} c_{\bi{i}}^\dag
(\bi{\sigma}\times
\bi{d}_{\bi{i}\bi{j}})_{z}c_{\bi{j}}+\sum_{\bi{i}}m_{i}c_{\bi{i}}^\dag
c_{\bi{i}} +g\sum\limits_{\bi{i}} c_{\bi{i}}^\dag \sigma_{z}
c_{\bi{i}} \ .\label{hamilton}
\end{eqnarray}
Here, the first term is the usual nearest neighbor hopping term with
$c_{\bi{i}}^\dag = ({c^\dag_{\bi{i},\uparrow }}
,{c^\dag_{\bi{i},\downarrow }} )$ as the electron creation operator
on site $\bi{i}$, where the hopping integral is set to be unity. The
second term is the intrinsic spin-orbit coupling (SOC) with coupling strength $v_{so}$, where
$\langle\langle \bi{i},\bi{j}\rangle\rangle$ stands for
the second nearest neighbor sites, and
$v_{ij}=(\bi{d}_{kj}\times \bi{d}_{ik})_{z}/|(\bi{d}_{kj}\times
\bi{d}_{ik})_{z}|$. Here, $\bi{k}$ is the common nearest neighbor of
$\bi{i}$ and $\bi{j}$, and vector $\bi{d}_{ik}$ points from $\bi{k}$
to $\bi{i}$. $v_{ij}$ = +1 for counter-clockwise
hopping, and $v_{ij}$  = -1 otherwise. $\sigma_{z}$ is a Pauli matrix describing the electron¡¯s spin.
The intrinsic SOC
opens a band gap and drives the system into the QSH phase.
 The third term stands for the  nearest neighbor Rashba SOC with
$\bi{\sigma}$ the Pauli matrix, and $v_{r}$ is Rashba SOC strength. The intrinsic SOC term breaks the $SU(2)$ symmetry down to $U(1)$,
and the Rashba SOC term  breaks the remaining $U(1)$ spin
symmetry down to $Z_{2}$ . The fourth term stands for a
staggered sublattice potential $(m_{i}=\pm m)$, which also opens a gap at the Dirac point but drives
the system into a trivial topology phase. The last term
represents a uniform exchange field with strength $g$,  which explicitly
violates the time reversal symmetry.

We consider systems with cylinder or torus boundary conditions,
consisting of $N_{x}$ ($N_{x}$ to be even) zigzag chains along the
circumferential direction ($y$ direction). The size of the sample
will be denoted as $N=N_{x}\times N_{y}$ with $N_{y}$ as the number
of atomic sites on each chain. We perform the entanglement cut along
the $y$ direction, which results in one or two interfaces between
the two equal parts $A$ and $B$, respectively, for the cylinder or
torus geometry, as shown in Fig.\ 1. In order to examine the EE and
ES, an Schmidt decomposition on the ground-state wavefunction or
calculation of the RDM is usually needed.  For non-interacting
fermion systems, however, the necessary information of the
entanglement can also be obtained from the following two-point
correlators~\cite{dmrg}
\begin{equation}\label{qw1}
c_{\tau_{1},\tau_{2}}(\bi{i},\bi{j}) = \langle
c^{\dagger}_{\bi{i},\tau_{1}}{c_{\bi{j},\tau_{2}}} \rangle \ .
\end{equation}
Here, $\langle \cdot \rangle $ means the ground-state expectation of
an operator. $\tau$ can be an index of spin, pseudospin, or orbital
degree of freedom.

Using the Fourier transformation (FT) along the $y$ direction, the
Hamiltonian can be rewritten as $H=\sum_{k_{y},i,j}c^\dag
_{i}(k_{y}) h_{i,j}(k_{y}) c_{j}(k_y)$, where $c^\dag
_{i}(k_{y})=(c^\dag _{i,\uparrow}(k_{y}),c^\dag
_{i,\downarrow}(k_{y}))$ are the electron creation operators. After
performing the entanglement cut, we treat part $A$ as the subsystem,
and trace out the degrees of freedom of $B$. It should be noted that
any of the correlators $c_{\tau_{1},\tau_{2}}(\bi{i},\bi{j})$ with
$\bi{i}$ and $\bi{j}$ confined in $A$  is unchanged by the tracing.
When carrying out the FT on the correlators, we can get
\begin{equation}
c_{\tau_{1},\tau_{2}}(\bi{i},\bi{j})=\frac{1}{N_{y}}\sum_{k_{y}}e^{ik_{y}\cdot(i_{y}-j_{y})}\langle
c^\dag_{i,\tau_{1}}(k_{y})c_{j,\tau_{2}}(k_{y}) \rangle   \ ,
\end{equation}
where $i$ and $j$  discriminate  the zigzag chains. We use $\langle
c^\dag_{i,\tau_{1}}(k_{y})c_{j,\tau_{2}}(k_{y}) \rangle$   to form a
hermitian matrix ${\cal C}(k_{y})$. Then the entanglement
Hamiltonian is given by ~\cite{dmrg}
\begin{equation}
    H_{ent}=\ln({\cal C}^{-1}-1)\ .
\end{equation}
The spectrum $\{\zeta_{i}\}$ of ${\cal C}$ is related to  spectrum
$\{\varepsilon_{i}\}$ of $H_{ent}$ by
$\zeta_{i}=1/(e^{\varepsilon_{i}}+1)$, where $\zeta_{i}$ acts as the
average fermion number in the entanglement energy level
$\varepsilon_{i}$ at ``temperature'' $T=1$. By using the spectrum of
${\cal C}$, the EE at each $k_{y}$ sector is given by
$s_{ent}(k_{y})=\sum_{i}s_{i}$, with
\begin{equation}
    s_{i}=-\zeta_{i}\ln\zeta_{i}-(1-\zeta_{i})\ln(1-\zeta_{i})     \ .
\end{equation}

\begin{figure}
\begin{center}
\includegraphics[width=3.2in]{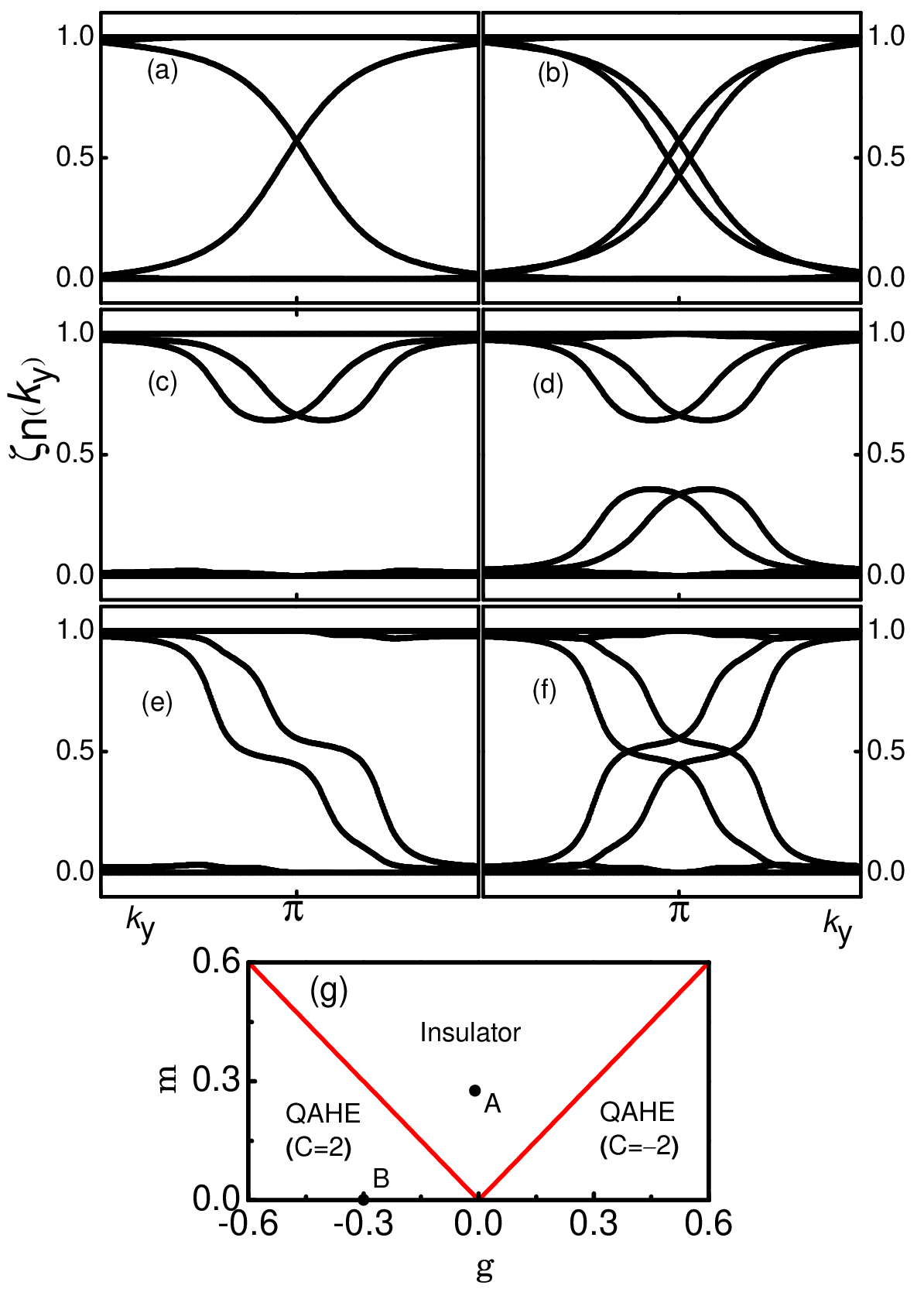}
\caption{ (a-f)  Single-particle entanglement spectrum in the cylinder geometry (left
panels) and torus geometry (right panels) for (a,b) the QSH phase with $v_{so}=m=0.2$, $v_{r}=0.1$, $g=0$, (c,d) the insulator phase with $v_{r}=-0.3$, $m=0.3$, $v_{so}=g=0$, and (e,f) the quantum anomalous Hall phase $v_{r}=g=-0.3$, $v_{so}=m=0$. (g) Phase diagram in the $m$ versus $g$ plane for
$v_{so}=0$ and $v_{r}\neq 0$. Points $A$ and $B$ correspond to the
parameter values used in (c,d) and (e,f), respectively.} \label{ES}
\end{center}
\end{figure}
From the viewpoint of probability theory, $s_{i}$ in  Eq.\ (6) can
be regarded as the Shannon (information) entropy of the Bernoulli
distribution, i.e., the $i$-th entanglement level $\varepsilon_{i}$
has probability $\zeta_{i}$ of being occupied while $(1-\zeta_{i})$
of being unoccupied. As a result, $S_{ent}$ is the Shannon entropy
of a series of such independent Bernoulli distributions. In the
following, we will perform systematic numerical simulations to study
various phases of Hamiltonian (\ref{hamilton}) in terms of the ES
and the subsystem particle number.

\section{ Entanglement spectrum and subsystem particle number}

At $g=0$, Hamiltonian (\ref{hamilton}) is the standard Kane-Mele
model~\cite{qshe1}, which is invariant under time reversal symmetry.
The system is in a QSH phase when
$|m/v_{so}|<[9-\frac{3}{4}(v_{r}/v_{so})^{2}]$, and is an insulator
when $|m/v_{so}|>[9-\frac{3}{4}(v_{r}/v_{so})^{2}]$. On the other
hand, if we set $v_{so}=0$, $v_{r}$ and $g$ nonzero, a middle band
gap opens when $|g|\neq|m|$. The system is in a quantum anomalous
Hall phase with Chern number $C=\pm 2$~\cite{yang} for $|g|<|m|$,
and is an insulator for $|g|>|m|$. The band gap closes at the
transition point $|g|=|m|$. The phase diagram for $v_{so}=0$ and
$v_{r}\neq 0$ is plotted in Fig.~\ref{ES}(g).

Figs.~\ref{ES}(a) and (b) show the ES for the QSH phase,
Figs.~\ref{ES}(c) and (d) for the insulator phase, and
Figs.~\ref{ES}(e) and (f) for the quantum anomalous Hall phase.
Here, it should be emphasized that the nontrivial topological phases
exhibit gapless ES [Figs.~\ref{ES}(a), (b), (e), and (f)],
corresponding to physical gapless edge modes, and this property is
named as the \emph{spectral flow}~\cite{est3}, which has been explained by Qi and his coworkers for coupled conformally
invariant subsystems with left- and right-moving particles  in all chiral topological systems~\cite{esc2}. However, the spectral
flow is broken for the topologically trivial phase
[Figs.~\ref{ES}(c) and (d)], which is also consistent with the
property of the correspondent edge states.

In a recent work~\cite{trace}, the authors proposed a new
characteristic quantity called the ``trace index'' to describe
topological invariants, which is defined through a subsystem
particle number operator $N_{A}(k_{y})=\sum_{i\in A}
c^\dag_{i,k_{y}}c_{i,k_{y}}$. The expectation of $N_{A}(k_{y})$ is
given by
\begin{eqnarray}\label{aver_na}
\langle N_{A}(k_{y})\rangle =\langle GS|\sum_{i\in A}
c^\dag_{i}(k_{y})c_{i}(k_{y}) |GS\rangle  =\mbox{Tr}{\cal C} \ .
\end{eqnarray} %

\begin{figure}
\begin{center}
\includegraphics[width=3.6in]{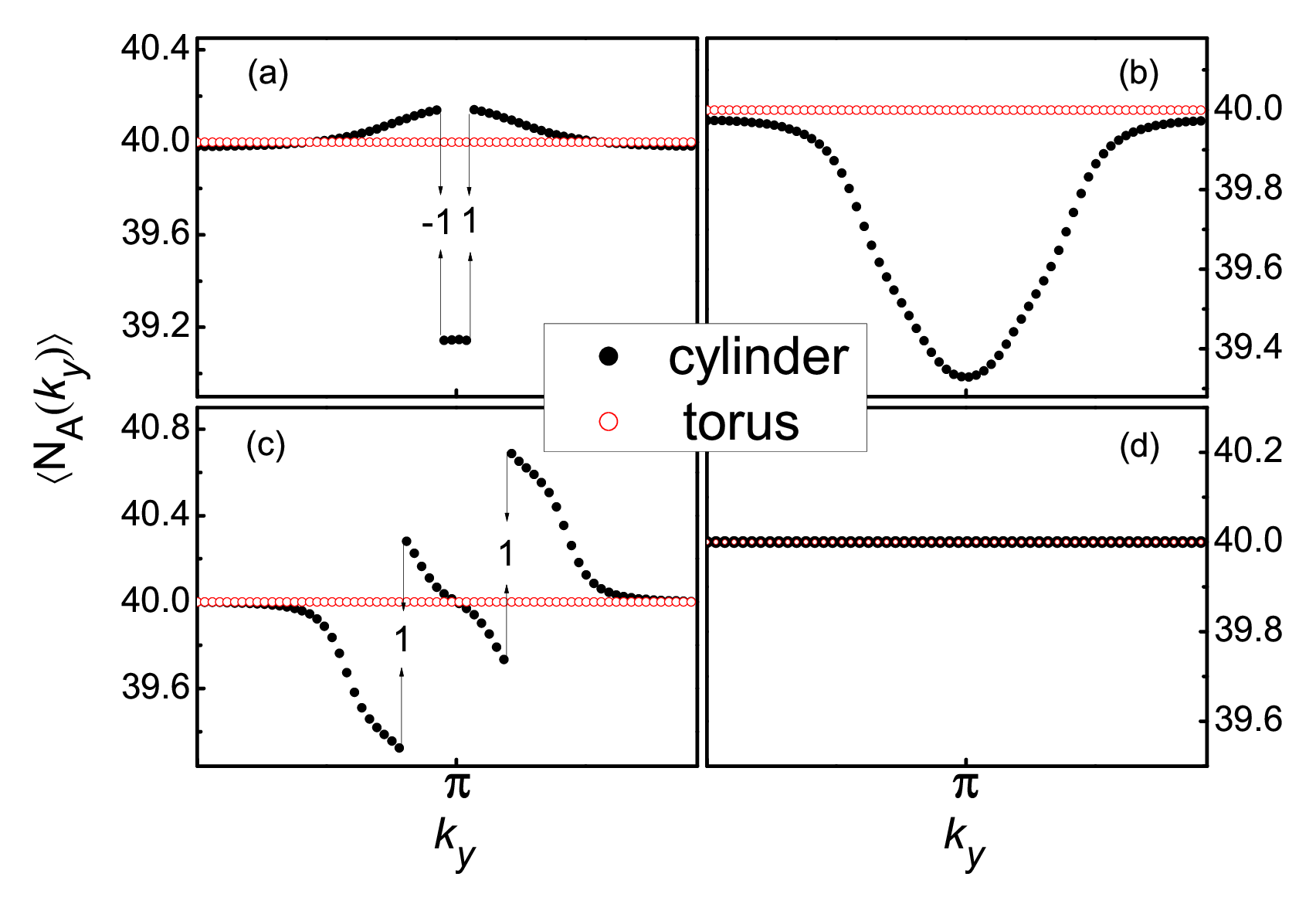}
\caption{(Color online) (a-c)The subsystem particle number in the
cylinder geometry and torus geometry for (a) the QSH phase, (b) the
insulator phase, and (c) the  quantum anomalous Hall phase with
all the parameters as same as Fig.\ 2. (d) The subsystem particle number in the
cylinder geometry and torus geometry for the QSH
phase with $v_{so}=v_{r}=0.2$, where the two-dimensional inversion
symmetry is retained. Discontinuities in the expectation of the
particle number can be observed only in the cylinder geometry. }
\label{trace1}
\end{center}
\end{figure}

In Fig.~\ref{trace1}, we plot the expectation of $N_{A}(k_{y})$ for
the three different phases mentioned above. In the cylinder
geometry, $N_{A}(k_{y})$ is discontinuous at some discrete momenta
in the nontrivial topological phases, as shown in
Figs.~\ref{trace1}(a) and (c). This is in contrast to the normal
insulator phase [see Fig.~\ref{trace1}(b)], where $N_{A}(k_{y})$ is a
continuous function of $k_{y}$. In the torus geometry,
$N_{A}(k_{y})$ is exactly equal to half of the total particle number
in the $k_y$ sector, without showing any discontinuity, because the
change of the particle number in $A$ around interface $I$ is just
canceled by that around interface $II$ due to the rotation
invariance of the torus. In the cylinder geometry, the \emph{trace
index} was defined as the total discontinuities of $\langle
N_{A}(k_{y})\rangle$ with varying momentum. Alexandradinata, Hughes,
and Bernevig~\cite{trace} presented a detailed analysis and proved
that the trace index is equivalent to the Chern number (or $Z_{2}$
invariant) for the Chern ($Z_{2}$) insulators. Therefore, the
subsystem particle number provides a new alternative tool to reveal
the topological invariants.

However, as mentioned in Ref.~\cite{trace}, there is an exceptional
case in which the subspace of the occupied bands at the symmetric
momenta is not closed under time reversal in the ground state. If at
the symmetric momenta the Kramers' doublet that extends along the
edge of $A$ is singly-occupied, $\langle N_{A}(k_{y})\rangle$ is
continuous, even when the system is in a nontrivial topological
phase. For the half-filled system  under consideration, an exception
still happens. While the two-dimensional inversion symmetry remains
unchanged ($m=0$), $N_{A}(k_{y})$ becomes continuous, as shown in
Fig.~\ref{trace1}(d). This is because  the Kramers' partners
extending along the edge simultaneously cross the Fermi level at the
symmetric momentum ($k_{y}=\pi$) and have opposite contributions to
the discontinuities of $\langle N_{A}(k_{y})\rangle$.

To overcome this difficulty,  we define a new quantity named spin
trace index.
 We choose operator $Ps_{z}P$ to split the fiber
bundle of the occupied states into two bundles with well-defined
Chern numbers, where $P$ is the ground state projector. At half
filling and in the presence of time reversal symmetry ($g=0$),
$Ps_{z}P$ is always a time-odd operator ($TPs_{z}PT^{-1}=-Ps_{z}P$),
so that the spectrum of $Ps_{z}P$ is symmetric in respect to the
origin. As a result, we can use  eigenvectors of operator $Ps_{z}P$
corresponding to the positive (negative) eigenvalues  to split the
Hilbert space spanned by the occupied-states wave functions  into
two sub-space ("spin-up" and "spin-down" sub-spaces).
 The new wave functions for the two sectors are unitary transformation of the original occupied-band
wave functions. This splitting results in a smooth decomposition
\begin{equation}\label{pro}
    P(k_{y})=P^{+}(k_{y})\oplus P^{-}(k_{y})           \ ,
\end{equation}
for all $k_{y}\in(0,2\pi]$, with $\alpha=\pm$ corresponding to the
positive and negative sectors. Straightforwardly, the two-point
correlator matrix can also be decomposed into ${\cal C}(k_{y})={\cal
C}^{+}(k_{y})\oplus {\cal C}^{-}(k_{y})$. It will be shown below
that the traces of ${\cal C}^{\pm}$, called the \emph{spin-projected
subsystem particle numbers}, are related to the topological
invariants.
\begin{figure}
\begin{center}
\includegraphics[width=4.0in]{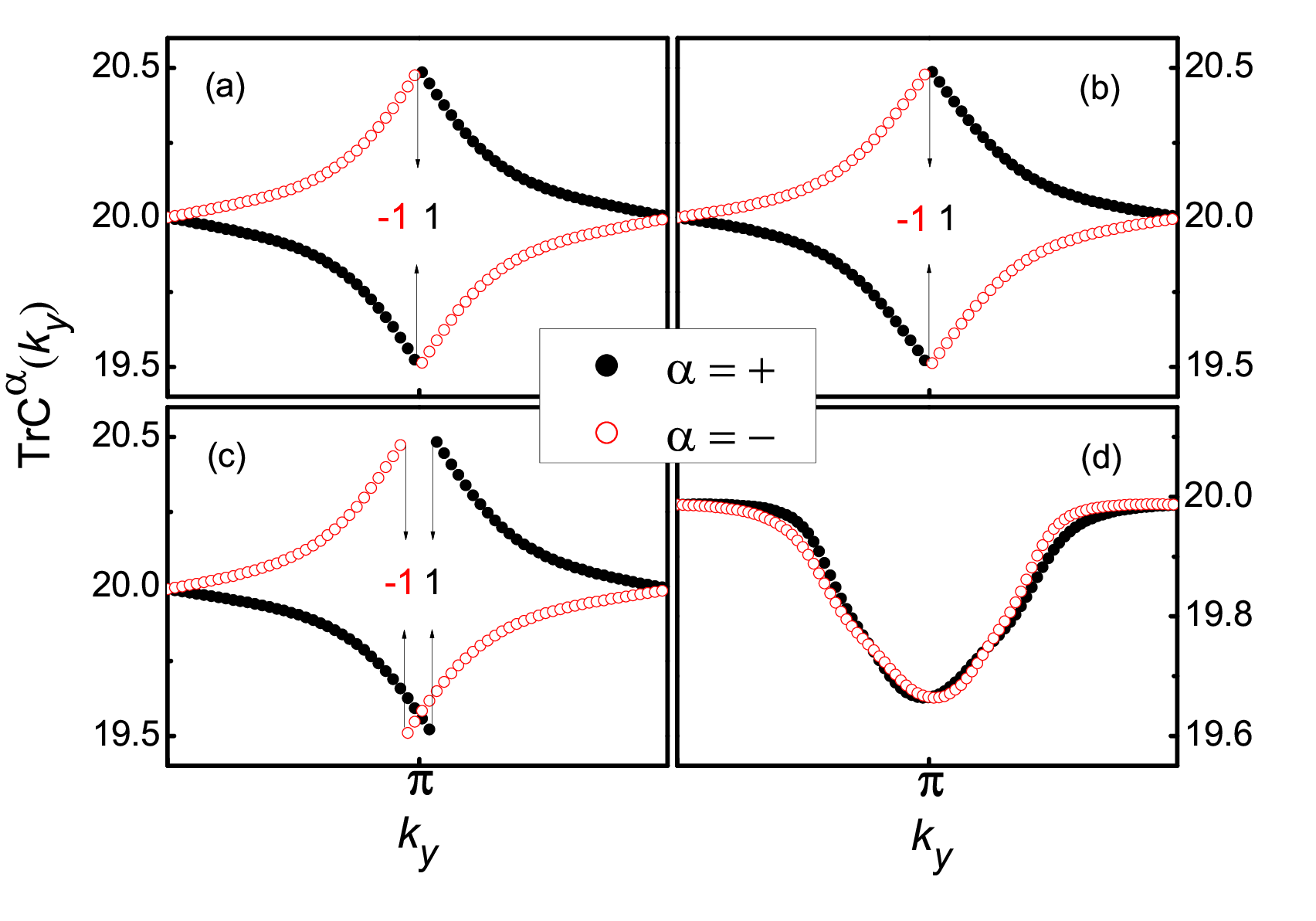}
\caption{(Color online) Subsystem spin-projected particle numbers in
the cylinder geometry for the QSH phase (a-c) with $g=0$ and
different parameters: (a) $v_{so}=0.2$ and $v_{r}=m=0$, (b)
$v_{so}=v_{r}=0.2$ and $m=0$, (c) $v_{so}=m=0.2$ and $v_{r}=0.1$,
and for the insulator phase (d) with $v_{so}=v_{r}=0.05$, $m=0.5$
and $g=0$.} \label{spin trace}
\end{center}
\end{figure}

If $\mbox{Tr}{\cal C}^{\alpha}(k_{y})$ is discontinuous at some
 momenta $\{k_{dis}\}$ with $k_{dis}\in (0,2\pi]$, we can
define the \emph{spin trace indices} as the total discontinuity,
i.e., difference between the limits of $ \mbox{Tr}{\cal
C}^{\alpha}(k_{y})$ from right and left in the thermodynamic limit,
\begin{equation}\label{A}
   A^{\alpha}\equiv\sum_{k_{dis}}(\lim_{k\rightarrow k_{dis+}}\mbox{Tr}{\cal C}^{\alpha}(k)-\lim_{k\rightarrow k_{dis-}}
   \mbox{Tr}{\cal C}^{\alpha}(k))                   \ .
\end{equation}

When the Hamiltonian commutes with $s_{z}$ ($v_{r}=0$), operator
$Ps_{z}P=I\otimes\sigma_{z}$. In this case, the spin trace index are
equivalent to the result obtained by calculating
$\sum_{i}<\varphi_{i}(k_{y})|(1\pm
\sigma_{z})/2|\varphi_{i}(k_{y})>$,  where $\varphi_{i}$ are limited
to the occupied-band wave functions. In the presence of the Rashba
spin-orbit coupling, the matrix form of  $Ps_{z}P$ is no longer
diagonal, and thence upper and lower projected spin spectral bands then begin
to communicate. Nevertheless, even in this case, it can be
confirmed that the two bands are always separated by a finite
band gap~\cite{spingap}, hence allowing us to define $A^{\alpha}$ unambiguously.
  Based on the splitting principle~\cite{split},   the
two sub-space got through  linear combination  still have
well-defined Chern number. It has been proved that the Chern numbers
for the two sectors are topological invariants protected by the
energy gap and spin spectrum gap~\cite{spinch2}. Then each sector is
an analogy of the quantum Hall system with Chern number $+1$ or
$-1$. The Chern number and trace index for a Chern insulator are
equivalent to each other~\cite{trace}. Naturally,  $A^{\alpha}$ is
equivalent to the Chern numbers  for each sector. From the quantized
Chern number for each sector, it follows that the spin trace index
is also quantized, both of them coming from the bulk topology. Then
$Z_{2}$ index can be defined as the parity of $A^{\alpha}$(for any
$\alpha$),
\begin{equation}\label{AZ2}
A_{Z_{2}}\equiv A^{\alpha} mod  \  2 ,
\end{equation}
which labels the topologically distinct phases.

We plot $\mbox{Tr}{\cal C}^{\alpha}$ $(\alpha=\pm)$ as functions of
$k_{y}$ in Fig.~\ref{spin trace}. No matter whether $s_{z}$ is
conserved, both $\mbox{Tr}{\cal C}^{+}(k_{y})$ and $ \mbox{Tr}{\cal
C}^{-}(k_{y})$ show discontinuities at  $k_{y}=\pi$ with $A^{+}=1$
and $A^{-}=-1$ in the QSH phase[Figs.~\ref{spin trace}(a) and (b)],
where the two-dimensional inversion symmetry is present ($m=0$).
Figure ~\ref{spin trace}(c) shows  the discontinuities of $
\mbox{Tr}{\cal C}^{+}(k_{y})$ and $ \mbox{Tr}{\cal C}^{-}(k_{y})$ in
the QSH phase in which $s_{z}$ is not conserved ($v_{r}\neq0$) and
the two-dimensional inversion symmetry is broken ($m\neq 0$). In
this case, the spin trace indices are  equal to 1 and $-1$,
respectively, contributed by two different momentum points. But, in
contrast, both $\mbox{Tr}{\cal C}^{+}(k_{y})$ and $ \mbox{Tr}{\cal
C}^{-}(k_{y})$ are the continuous functions in the normal insulator
phase [Fig.~\ref{spin trace}(d)].
 Consequently, it is easy to get $A_{Z_{2}}=1$ for QSH phase [Figs.~\ref{spin trace}(a-c)] and $A_{Z_{2}}=0$ for insulator phase [Figs.~\ref{spin trace}(d)]. Therefore, the subsystem particle number
expectation can be used to characterize the topological invariants.
Especially, for the QSH systems, the spin trace indices are new
well-defined quantities that can reveal the $Z_{2}$ invariant and
distinguish different quantum phases.

\section{ Application of spin trace indices to disordered system}
\begin{figure}
\includegraphics[width=4.4in]{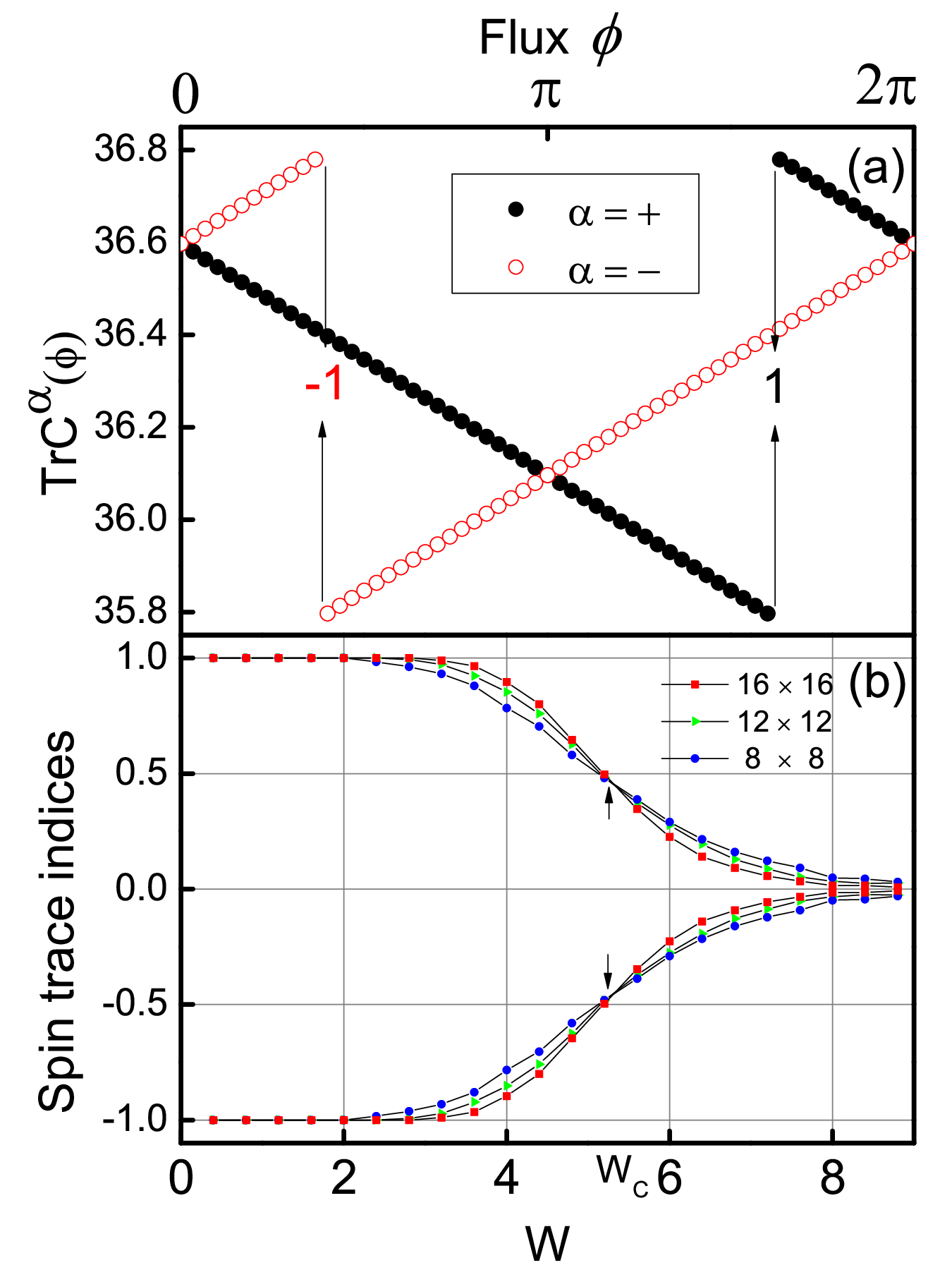}
 \centering
\caption{(Color online) Disordered spin trace indices for the QSH phase with $v_{so}=m=0.2$ and $v_{r}=0.1$ .
(a): $\mbox{Tr}{\cal C}^{\alpha}(\phi)$  as a function of flux $\phi$ for one time disorder configuration with disorder strength parameter $W=0.5$ for sample size $N = 12 \times 12$;
 (b)Disordered trace indices  as a function of the disorder strength parameter
$W$. The result is averaged over 400
disorder configurations, for several different sample sizes. }
\label{trace}
\end{figure}
The spin trace indices can be used for classifying different topological phases, but it will be equally important
to extract nontrivial physical consequences from its remarkable properties, in particular, to say
something about the localization of the bulk states in the presence of disorder. To study the disorder effect,
we include into  Hamiltonian (\ref{hamilton}) a random on-site potential of the
form $\sum_{\bm{i}}\omega_{\bm{i}}c^{\dag}_{\bm{i}}c_{\bm{i}}$, with $w_{\bm{i}}$ randomly distributed between
$[-W/2,W/2]$.

We define spin trace indices for systems with
translational invariance in one dimension, but momentum $k_{y}$ is no longer a good quantum number  in
the presence of disorder. However, by introducing a flux inserted through the symmetry
axis of the cylinder, we add a parameter $\phi$ into the Hamiltonian
in order to replace the momentum $k_{y}$. Then if
 $\mbox{Tr}{\cal C}^{\alpha}(\phi)$ is discontinuous at some
 pseudo-momenta $\{\phi_{dis}\}$ with $\phi_{dis}\in (0,2\pi]$, we can
redefine the disordered spin trace indices as:
\begin{equation}\label{AA}
   A_{\phi}^{\alpha}\equiv\sum_{\phi_{dis}}(\lim_{\phi\rightarrow \phi_{dis+}}\mbox{Tr}{\cal C}^{\alpha}(\phi)-\lim_{\phi\rightarrow \phi_{dis-}}
   \mbox{Tr}{\cal C}^{\alpha}(\phi))                   \ .
\end{equation}

Numerical calculation of the disordered spin trace indices is carried out for a
disordered system with $v_{so}=m=0.2$ and $v_{r}=0.1$ at half filling, and the
results are shown in Fig.\ \ref{trace}. In Fig.\ref{trace}(a), we plot $\mbox{Tr}{\cal C}^{\alpha}(\phi)$  as
a function of flux $\phi$ for one time disorder configuration with weak disorder $W=0.5$,  for sample size $N = 12 \times 12$.
It is apparent that for weak disorder, the $\mbox{Tr}{\cal C}^{\alpha}$ $(\alpha=\pm)$ as a function of
$\phi$ behave very similar to the $\mbox{Tr}{\cal C}^{\alpha}$ $(\alpha=\pm)$
as a function of $k_{y}$ which have been showed in Fig.~\ref{spin trace}. For
the QSH, both $\mbox{Tr}{\cal C}^{\alpha}$ have only one discontinuous momentum or pseudo-momentum point with discontinuity
$\pm 1$, which implies
the spin trace indices are  equal to 1 and $-1$,
respectively. When the disorder strength $W$ is
increasing,  $\mbox{Tr}{\cal C}^{\alpha}(\phi)$ become
discontinuous at some more pseudo-momenta $\{\phi_{dis}\}$ points and
the calculated spin trace indices as a function of disorder strength after
average over 400 times disorder configurations  is plotted in Fig.\
\ref{trace}(b) with different sample sizes. One can clearly found that the
spin trace indices are robust against weak disorder $W<2$. With increasing
$W$ from 2 to 9, the spin trace indices continuously decrease to nearly
zero. At the same time, with increasing the sample size, the transition process
becomes sharper and sharper, which conforms the expectation that the
phase transition
from the QSH (with spin trace indices $\pm 1$) to the trivial
insulator (with both spin trace indices 0) occurring at around $Wc \approx 5.2$  should become a sudden drop from $\pm 1$ to 0 in the
thermodynamic limit. In short, as far as disorder goes, loss of translational invariance leads to a loss of the Brillouin zone, and sequentially
the $Z_{2}$ index can not be defined, so how to abstract the topological invariant from disordered systems becomes very significant. Here the (disordered) spin trace indices we has defined above can be applied to these systems and can display a topological phase transition occuring when the
disorder strength is increased beyond a critical values.

\section{ Entanglement entropy and subsystem particle number fluctuation}
We have shown that topological properties of the ground state can be
extracted from the expectation of subsystem particle number. Now we turn to the variance of
$N_{A}(k_{y})$. In the past three years, extensive works have been devoted
to the study of the relation between the EE and subsystem particle fluctuation for
non-topological systems~\cite{fluc1}. In this section, we will show that
the relation is rather general, it does apply to non-interacting electron systems
with a nontrivial band topology.
We start from the standard definition of the variance
\begin{eqnarray}\label{var1}
 \triangle N^{2}_{A}(k_{y}) =\langle  N^{2}_{A}(k_{y}) \rangle-\langle N_{A}(k_{y})\rangle^{2}
\ .
\end{eqnarray}
Substituting  Eq.~(\ref{aver_na}) into Eq.~(\ref{var1}) and using
the Wick's theorem to expand all the four-point correlators, one can
obtain
\begin{eqnarray}\label{var2}
\triangle N^{2}_{A}(k_{y}) &= \sum_{i,j\in A} \langle
c^\dag_{i,k_{y}}c_{j,k_{y}}\rangle \langle
c_{j,k_{y}}c^\dag_{i,k_{y}}\rangle \nonumber\\ &=\mbox{Tr}[{\cal
C}(1-{\cal C})]          \ ,
\end{eqnarray}
yielding $\triangle
N^{2}_{A}(k_{y})=\sum_{i}\zeta_{i}(1-\zeta_{i})$, which is in
keeping with the variance formula of the Bernoulli distributions.
\begin{figure}
\begin{center}
\includegraphics[width=3.6in]{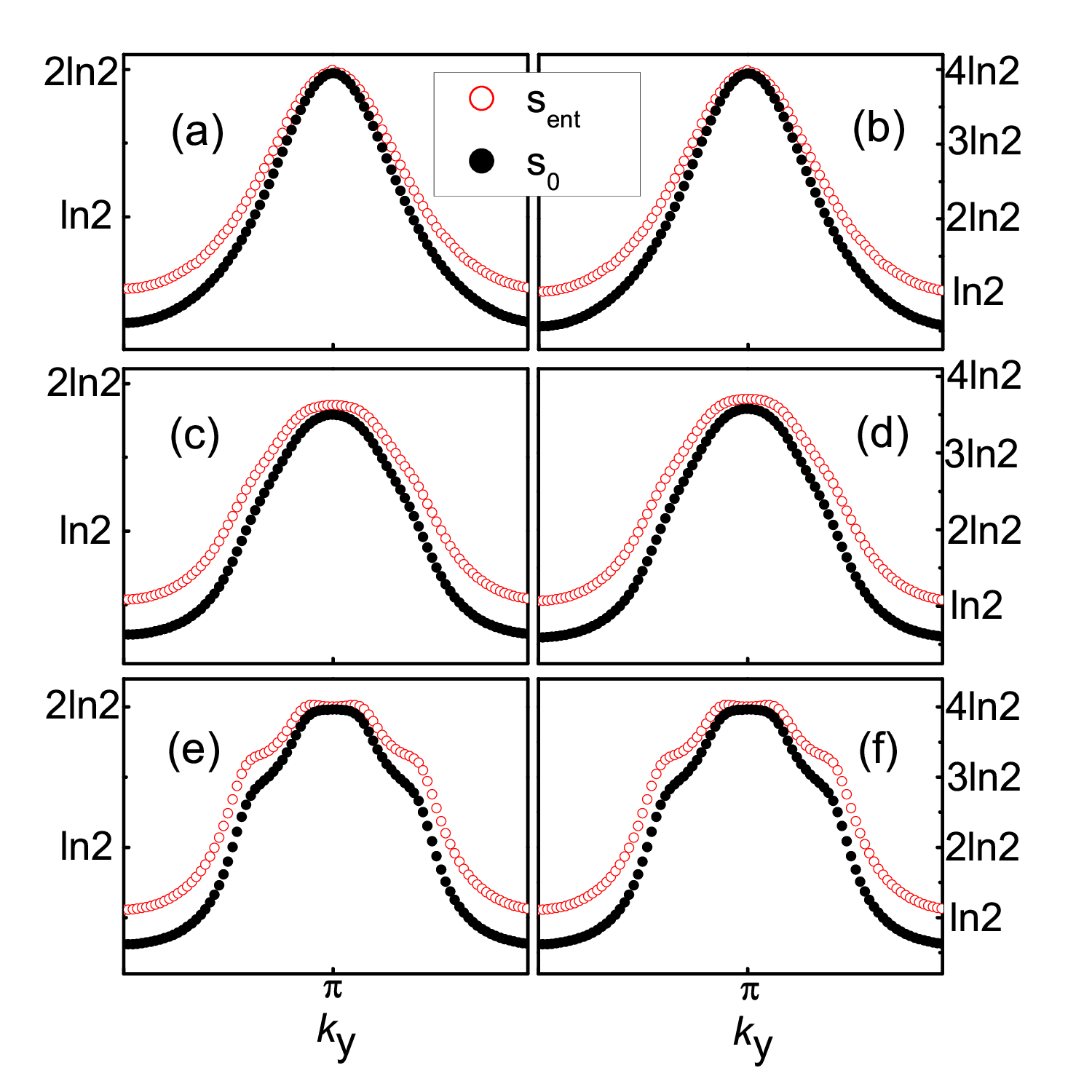}
\caption{(Color online) Entanglement entropy in comparison with
subsystem particle number fluctuation for (a-b) the QSH phase,
 (c-d) the insulator phase, and (e-f) the quantum anomalous Hall
phase, in the cylinder geometry (left panels) and torus
geometry (right panels). All the parameters are the same as in
Fig.\ 2. } \label{ee}
\end{center}
\end{figure}
In order to find a definite relationship between the EE and the
variance, one can construct a concave function $f(x)=-\ln x/(1-x)$
for $x\in[0,1]$, and  apply the Jensen's inequality
\begin{equation}\label{inequ}
    -x\ln x-(1-x)\ln(1-x)  \geq (4\ln 2)\cdot x(1-x)\ .
\end{equation}
The equality holds if and only if $x=1/2$. Equation (\ref{inequ})
enables us to make a lower-bound estimation of the EE
\begin{equation}
    s_{ent}(k_{y}) \geq (4\ln 2)\cdot \triangle N^{2}_{A}(k_{y})     \ .
\end{equation}

This inequality is first given in the context of metal~\cite{Klich}
and here as a complement, we give a very simple and  direct proof.
From the inequality  one can see that a lower bound of the EE is given by
$s_{0}(k_{y})\equiv(4\ln 2)\cdot \triangle N^{2}_{A}(k_{y})$,
which is directly proportional to the particle number fluctuation of
subsystem. In Fig.~\ref{ee}, we plot $s_{ent}(k_{y})$ and
$s_{0}(k_{y})$ in the QSH phase, insulator phase, and quantum
anomalous Hall phase. In all the cases, the curves for the particle
number fluctuation behave somewhat similarly, and are very close to
the corresponding EE. This similarity was observed
in the non-topological systems lately~\cite{fluc1}, and
here we find that the similarity remains to hold for
the topologically nontrivial system.
Moreover, each maximally entangled state with
$\varepsilon_{m}=0$ $(\zeta_{m}=1/2)$ existing only in topology  phases contributes a maximal value to
the subsystem particle number fluctuation and the EE, which cannot be eliminated by adiabatic continuous
deformation, which may provides a new way for the probe of topological insulators.

Furthermore, one can use
$N_{A}(k_{y})=\sum_{i\in A} c^\dag_{i,k_{y}}c_{i,k_{y}}$ to verify
$\triangle N^{2}_{A}=\sum_{k_{y}}\triangle N^{2}_{A}(k_{y})\rightarrow\frac{L_{y}}{2\pi}\int dk_{y}\triangle N^{2}_{A}(k_{y})$, indicating that
$\triangle N^{2}_{A}(k_{y})$ satisfies a  area law~\cite{area},
similar to the EE, $S_{ent}=\sum_{k_{y}}s_{ent}(k_{y})\rightarrow
\frac{L_{y}}{2\pi}\int dk_{y}s_{ent}(k_{y})$. Remarkably, for topologically-ordered states, Ref~\cite{geoent} proved that the non-topologically-ordered term of geometric entanglement obeys a similar area law in  multipartite entanglement. We expect that the conclusion should still be true for non-interacting fermions systems.
To conclude, the
subsystem particle number fluctuation shares several common
characteristics with the EE, and so can be utilized to obtain an
estimate of the EE. The EE has being widely used in analyzing
quantum critical phenomena, topologically ordered states, evolution after a quantum quench,
as well as quantum computation~\cite{eeapl}.

\section{ Summary and discussion}
To conclude, we have investigated the relationship between the
quantum entanglement and subsystem particle number. The spin trace
indices can reveal the topological invariants and be used to
classify different phases in QSH systems. This new tool always works
well even though $s_{z}$ is not conserved. Even in disordered system,
it works well and can be used to demonstrate topological phase transition.
As to the subsystem particle number fluctuation, it shares several common properties
with the EE. They both satisfy the same area law, and are dominated
by the boundary excitations with each zero mode having a maximal
contribution. The connection between the two quantities is
universal, regardless of whether the system has a nontrivial band
topology. As a result, the subsystem particle number fluctuation, as
an observable quantity, can be utilized to obtain an
estimate of the EE experimentally~\cite{fluc1}.

As long as the Fermi energy still lies in the bulk energy gap,
 all results that we have obtained will remain about the same.
With the Fermi energy lowered into
 the valence band or increased
into the conduction band, the spin trace indices will no longer be quantized and continuously
 drop from $\pm1$ to 0.  We also stress that the results obtained in the paper only hold for free fermion systems.
Interestingly, as to the entanglement entropy and the subsystem particle number fluctuation, a similar relation has been found for certain types of interacting systems, for example, one dimension quantum spin
chains~\cite{fluc2}, although such a relation is not true for some other interacting systems such as fractional quantum Hall states~\cite{fluc3} and two-dimensional spin $1/2$ antiferromagnetic Heisenberg model~\cite{fluc4}.

\ack
This work was supported by the State Key Program for Basic Researches of China under
grants numbers 2014CB921103 (LS), 2011CB922103 and 2010CB923400 (DYX), the National
Natural Science Foundation of China under grant numbers 11225420 (LS), 11304281(RW),
11174125, 91021003 (DYX) and a project funded by the PAPD of Jiangsu Higher
Education Institutions.

\end{document}